\documentclass[conference]{IEEEtran}
\IEEEoverridecommandlockouts
\usepackage{cite}
\usepackage{flushend}
\usepackage{amsmath,amssymb,amsfonts}
\usepackage{graphicx}
\usepackage{textcomp}
\usepackage{xcolor}
\usepackage[utf8]{inputenc}
\usepackage[T1]{fontenc}
\usepackage{algorithm}
\usepackage{algorithmic}
\usepackage{comment}
\def\BibTeX{{\rm B\kern-.05em{\sc i\kern-.025em b}\kern-.08em
    T\kern-.1667em\lower.7ex\hbox{E}\kern-.125emX}}
\usepackage[normalem]{ulem}
\usepackage{pifont}
\usepackage{subcaption}
\usepackage{array}
\usepackage{caption}
\usepackage{booktabs}
\usepackage{multirow}
\usepackage{hyperref}
\hypersetup{colorlinks=true, linkcolor=blue, filecolor=magenta, urlcolor=cyan, citecolor=blue}

\def\BibTeX{{\rm B\kern-.05em{\sc i\kern-.025em b}\kern-.08em
    T\kern-.1667em\lower.7ex\hbox{E}\kern-.125emX}}
\begin{document}

\title{Explainability Boosted Anomaly Detection Framework for O-RAN based NextG Networks}

 \author{
    \IEEEauthorblockN{Nurullah Aksu\IEEEauthorrefmark{1}, Ali Fuat Sahin\IEEEauthorrefmark{1}{\IEEEauthorrefmark{4}}, Semiha Tedik Başaran\IEEEauthorrefmark{1}}
    \IEEEauthorblockA{\IEEEauthorrefmark{1}Faculty of Electrical and Electronics Engineering, Istanbul Technical University, Istanbul, Turkiye}
    \IEEEauthorblockA{\IEEEauthorrefmark{4}Communications and Signal Processing Research (HISAR) Lab, TUBITAK BILGEM, Kocaeli, Turkiye}
    \IEEEauthorblockA{Email: \{aksun19, sahinal18, tedik\}@itu.edu.tr}}   

\maketitle

\begin{abstract}
The wireless networks have historically faced significant security vulnerabilities, necessitating advanced anomaly detection mechanisms, especially as networks evolve towards 6G and beyond. This study introduces an advanced anomaly detection framework that leverages explainable artificial intelligence to enhance the security of next-generation (NextG) cellular networks. By implementing and evaluating a variety of artificial intelligence models, the framework demonstrates high accuracy and efficient runtime performance in identifying malicious traffic within a realistic Open Radio Access Network (O-RAN) testbed. A key innovation of this work is the integration of post-hoc explainability methods to identify the most critical key performance metrics (KPMs), which enables a significant 80\% reduction in dataset complexity without compromising detection accuracy. Additionally, explainability analyses identify several critical attack traffic characteristics, such as protocol type, bandwidth, interval, and duration, to prevent upcoming network attacks. The resulting framework effectively balances computational efficiency, accuracy, and explainability, underscoring its practical applicability for enhancing security in next-generation cellular networks.
\end{abstract}


\section{Introduction} 
As networks have historically been vulnerable to a variety of security threats, the security of cellular communication systems remains critically important. Prior generations exhibited exploitable weaknesses despite various developments, and today’s security challenges are expanded by evolving attack methods, rapid user growth, and increasingly complex architectures. In response, next-generation (NextG) networks introduce stronger, standardized security mechanisms. Nonetheless, adversaries continue to adapt, with attackers employing diverse strategies aimed at disrupting communication, stealing sensitive information, and causing service degradation, which can be classified as "anomalies" such as signalling storms and volumetric attacks \cite{attacks}. Given these persistent security challenges, enhancing the security mechanisms of NextG networks has emerged as a significant area of research. Recent literature highlights multiple approaches to developing effective anomaly detection algorithms \cite{fractal, utku_stats}. The anomaly detection approach in \cite{fractal} relies on fractal analysis of time-series data, whereas the method in \cite{utku_stats} examines the characteristics of 3GPP radio protocols using statistical and ML techniques rather than analyzing conventional IP packets.

While these methods have demonstrated promising results, their integration into conventional cellular network infrastructures remains a significant challenge. To address this issue, the Open Radio Access Network (O-RAN) concept has emerged \cite{o-ran}. O-RAN promotes a flexible, disaggregated, and vendor-agnostic network architecture, contrasting sharply with traditional approaches. The key benefits of O-RAN are the elimination of vendor lock-in and the facilitation of distributed contributions \cite{sahin2023oai}. A particularly significant component proposed by the O-RAN Alliance is the RAN Intelligent Controller (RIC), a software-defined network element designed to manage control functionalities through custom external applications known as xApps in near-real-time (near-RT) \cite{o-ran}.

Recent studies \cite{access_anomaly_xapp, ric_ddos, srsRAN_ric, autoencoder} have utilized the flexibility offered by O-RAN architectures to develop various xApps tailored for anomaly detection and mitigation purposes. For instance, a threshold-based anomaly detection algorithm implemented via an xApp has been proposed in \cite{access_anomaly_xapp}. Similarly, \cite{ric_ddos, srsRAN_ric} present ML-based xApps designed to detect signalling storm attacks, while an autoencoder-based model is designed in \cite{autoencoder} to leverage Artificial Intelligence (AI) models for anomaly detection. In addition to O-RAN research, \cite{federated_oran} has utilized the Colosseum platform to collect benign and malicious time-series traffic to train a federated learning-based anomaly detection model. While the data in \cite{federated_oran} is not accessible, public datasets featuring Denial-of-Service attack scenarios in \cite{KDDCup1999, NSL-KDD1999, nidd} have been long studied. \cite{KDDCup1999} includes both benign and malicious connection data, representing a standard benchmark set with various simulated intrusions. The refined version of \cite{KDDCup1999} is \cite{NSL-KDD1999}, which refines the dataset to reduce evaluation bias. In contrast, \cite{nidd} is generated from a real-time 5G test network to enhance the reliability even further.  

\begin{figure*}[t]
    \centering
    \includegraphics[width=\textwidth]{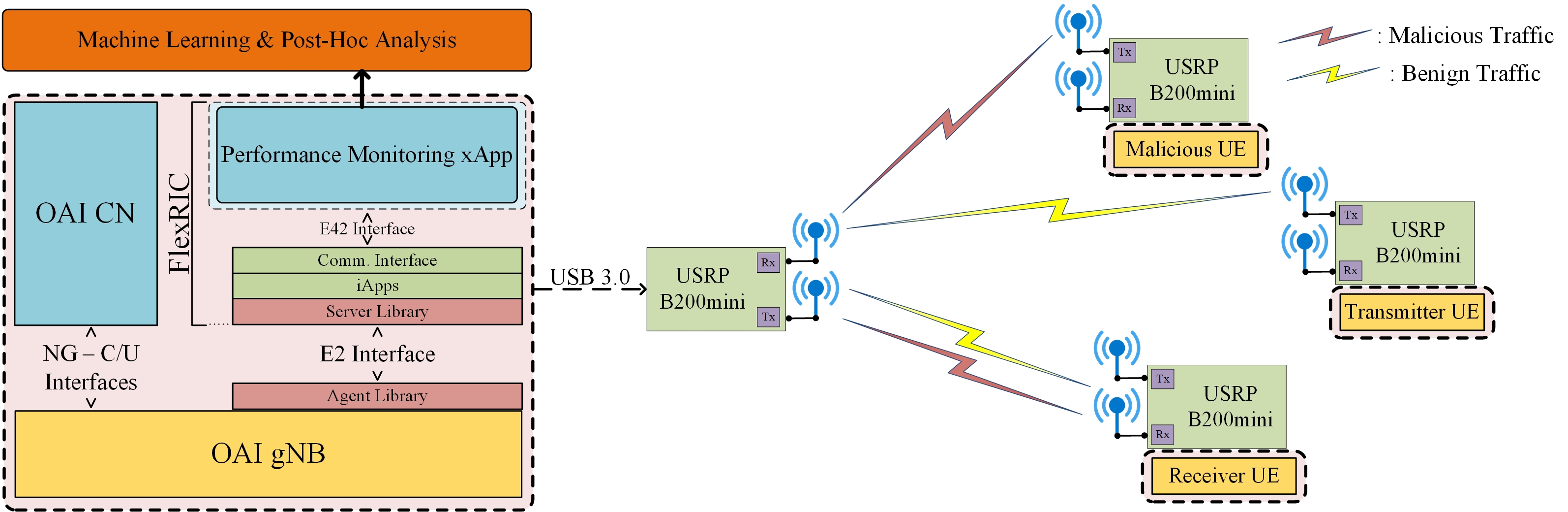}
    \caption{Attack detection scenario within an O-RAN network with legitimate and malicious UEs.}
    \vspace{-15pt}
    \label{fig:systemModel}
\end{figure*}
\label{sysMod}

Motivated by these security needs, this study proposes a near-RT RIC-based anomaly detection framework employing explainability-boosted AI/ML to strengthen security in NextG networks. First, benign and malicious traffic scenarios are generated realistically, and a monitoring xApp is developed to record detailed characteristics of the 3GPP radio protocol stack. Using the constructed dataset, multiple algorithms are trained and evaluated. Subsequently, statistical-based and post-hoc methods are implemented to identify and extract the most influential features for anomaly detection. Lastly, explainability analyses identify the critical attack characteristic to prevent upcoming network attacks.

\section{System Model}
\vspace{-3pt}
This study employs multiple software and hardware components to establish and analyze a NextG communication network for anomaly detection and analysis. Traffic data are obtained from an open-source real-time 5G communication system, an open-source near-RT RIC, dedicated traffic generators (i.e., hping and iperf3), and a monitoring xApp. These tools collectively produce both benign and malicious traffic, while the xApp collects key performance metrics from the near-RT RIC platform for subsequent anomaly detection and explainability analysis. This integrated setup enables end-to-end evaluation of anomaly detection techniques together with explainability methods under realistic NextG traffic conditions. 

Fig. \ref{fig:systemModel} illustrates the overall real-time system architecture, featuring 5G core network (CN), one gNodeB (gNB), and three user equipments (UEs). All UEs are properly registered to the network through the authentication in the CN. While two UEs generate legitimate network traffic, the remaining UE is designated to generate malicious traffic aimed at one of the legitimate users in the network, thereby creating an attack scenario. The legitimate UEs generate traffic patterns consistent with typical 5G applications, while the malicious UE performs sophisticated attack strategies to degrade the service quality of the legitimate UEs. In Fig. \ref{fig:systemModel}, the benign traffic is shown with the yellow, whereas the malicious traffic is shown with the red. By injecting both benign and malicious traffic flows concurrently, more challenging conditions can be tested by AI/ML-based anomaly detection models.

The 5G communication system in this study is implemented using the open-source OpenAirInterface (OAI) platform \cite{OAI}, coupled with FlexRIC \cite{flexric} to provide near-RT RIC functionality alongside a monolithic 5G base station. For network settings, the frequency band n78 ($3.6$ GHz) with 106 PRBs (20 MHz bandwidth) and subcarrier spacing of 30 kHz have been used. Additionally, each network element employs a software-defined radio as its radio-frequency frontend, specifically the USRP B200mini model. Lastly, the iperf3 command has been exploited to emulate benign traffic, whereas the hping3 command has been utilized to produce malicious traffic.

\subsection{OpenAirInterface \& FlexRIC} 
The limited number of network providers can impede the development of next-generation mobile communications. In response, O-RAN promotes open-source collaboration by decoupling the network development from specific vendors, thereby accelerating innovation. Among these initiatives, OAI stands out as a prominent open-source platform that supports critical end-to-end 5G functionalities. The OAI codebase is continuously improved to satisfy industry standards through ongoing research projects, ensuring its alignment with the latest trends and advancements in the field.
Using OAI, an end-to-end communication system that includes a core network, a gNB, and three UEs has been established. This setup allows researchers to rapidly prototype, modify, and evaluate new 5G features under realistic conditions. 

FlexRIC, similar to the OAI, is an open-source platform that provides a near-RT RIC solution enabling custom logic integration into 5G networks without significant structural modifications. This flexible architecture supports the development of xApps, which can analyze or manipulate network behavior in near-RT. Consequently, any new control or analytics function can be introduced swiftly. In this study, FlexRIC is deployed to monitor user metrics collected by the base station.

\begin{table*}[htbp]
\footnotesize
\centering
\vspace{10pt}
\caption{Metrics Description}
\vspace{-5pt}
\label{kpmMetrics}
\begin{tabular}{|c|p{0.38\columnwidth}|p{0.46\columnwidth}|p{0.08\columnwidth}|p{0.85\columnwidth}|}
\hline
\multicolumn{1}{|c|}{\textbf{\#}} & 
\multicolumn{1}{c|}{\textbf{Metric Name}} & 
\multicolumn{1}{c|}{\textbf{Description}} & 
\multicolumn{1}{c|}{\textbf{Unit}} & 
\multicolumn{1}{c|}{\textbf{Usage}} \\ \hline
1 & \textbf{\ding{72}} DRB.PdcpSduVolumeDL  & PDCP SDU volume in the downlink  & kb & Monitors the downlink data transmitted at the PDCP layer. \\ \hline
2 & \textbf{\ding{72}} DRB.PdcpSduVolumeUL  & PDCP SDU volume in the uplink    & kb & Monitors the uplink data transmitted at the PDCP layer. \\ \hline
3 & \textbf{\ding{72}} DRB.RlcSduDelayDL    & RLC SDU delay in the downlink    & $\mu$s & Measures latency at the RLC layer in the downlink. \\ \hline
4 & \textbf{\ding{72}} DRB.UEThpDL          & UE throughput in the downlink    & kbps & Monitors the data rate in the downlink. \\ \hline
5 & \textbf{\ding{72}} DRB.UEThpUL          & UE throughput in the uplink      & kbps & Monitors the data rate in the uplink. \\ \hline
6 & DRB.UEBufferSizeUL   & UE buffer size in the uplink     & kb & Indicates the uplink buffer occupancy at the UE. \\ \hline
7 & DRB.UETotPduDL       & Total PDUs in the downlink       & — & Tracks the total number of PDUs transmitted in downlink. \\ \hline
8 & DRB.RlcBytesInBuffer & Data volume stored in RLC        & kb & Indicates how much data at the RLC layer is awaiting transmission. \\ \hline
9 & DRB.RlcPdusInBuffer  & Number of PDUs stored in RLC      & — & Indicates how many RLC PDUs are pending for transmission. \\ \hline
10 & \textbf{\ding{72}} RRU.PrbTotDL        & Total PRBs allocated in DL       & PRBs & Monitors the number of resource blocks allocated for downlink. \\ \hline
11 & \textbf{\ding{72}} RRU.PrbTotUL        & Total PRBs allocated in UL       & PRBs & Monitors the number of resource blocks allocated for uplink. \\ \hline
12 & RRU.PrbRetDL        & Retransmitted PRBs in DL         & PRBs & Counts the number of retransmitted PRBs in downlink. \\ \hline
13 & RRU.PrbRetUL        & Retransmitted PRBs in UL         & PRBs & Counts the number of retransmitted PRBs in uplink. \\ \hline
14 & MAC.BlerDL          & DL Block Error Rate              & — & Indicates block error rate in the downlink MAC layer. \\ \hline
15 & MAC.BlerUL          & UL Block Error Rate              & — & Indicates block error rate in the uplink MAC layer. \\ \hline
16 & TB.ErrTotalNbrDL    & Erroneous DL transport blocks    & — & Presents the reliability of downlink transport block transmission. \\ \hline
17 & TB.ErrTotalNbrUL    & Erroneous UL transport blocks    & — & Presents the reliability of uplink transport block transmission. \\ \hline
18 & TB.HARQEffDL        & HARQ efficiency in DL            & \% & Measures downlink HARQ success rate. \\ \hline
19 & TB.HARQEffUL        & HARQ efficiency in UL            & \% & Measures uplink HARQ success rate. \\ \hline
20 & CARR.PUSCH-MCS      & MCS index used for PUSCH         & — & Reflects the modulation and coding scheme used for uplink. \\ \hline
21 & CARR.PDSCH-MCS      & MCS index used for PDSCH         & — & Reflects the modulation and coding scheme used for downlink. \\ \hline
22 & L1M.SS-RSRP         & Reference Signal Received Power  & dBm & Measures signal strength from the synchronization signal block. \\ \hline
23 & L1M.PUSCH-SNR       & Signal-to-Noise Ratio of PUSCH   & dB & Evaluates the uplink signal quality of PUSCH transmissions. \\ \hline
\end{tabular}
\vspace{-10pt}
\end{table*}

\section{Data Collection Process}
\label{SectionIII}
To facilitate data collection, the ETSI Technical Specification 128.552 (Key Performance Metrics, KPMs) \cite{kpm} is adopted as a guideline within FlexRIC. The KPMs used in this study are listed in Table \ref{kpmMetrics}, where standardized metrics are marked with a star (i.e., \ding{72}), and additional metrics have been introduced to enhance the performance of the AI/ML models. Here, the selection of additional metrics was driven by the aim of covering across various protocol layers and attack archetypes. Some metrics capture spectrum occupancy and data transmission efficiency at the physical layer, while others reflect queue dynamics in upper layers, jointly characterizing radio-resource usage, link quality and fading conditions. This layered approach provides comprehensive observability, enabling discrimination among attack types through diverse network indicators.

Two primary traffic cases are considered for data collection on the physical testbed, as shown in Fig. \ref{fig:systemModel}: benign traffic and malicious traffic. Benign traffic is generated between a transmitting UE and a receiving UE. To simulate diverse applications, five different traffic patterns are selected: Web Browsing, Multimedia Streaming, File Transfer, Remote Access, and Instant Messaging to emulate the most common transmission protocols. Two bash scripts have been developed to transmit and receive benign traffic. The transmission durations and traffic patterns are predetermined, while the transmission rate is determined randomly in each loop for benign traffic. After all traffic flows are finished, another loop with random bandwidths is executed. The details of the benign traffic are provided in Table \ref{trafficPatterns}. Due to the constraints imposed by the iperf3 tool, a one-second timestep is employed during the data collection process.

The second case involves malicious traffic, generated by an attacker UE using three different transmission protocols: User Datagram Protocol (UDP), Transmission Control Protocol (TCP), and Internet Control Message Protocol (ICMP). These protocols enable four distinct signalling flood attacks to be performed: SYN, UDP, TCP, and ICMP, which are the most preferred attack types in the literature \cite{nidd,NSL-KDD1999}. Similar to benign traffic, malicious traffic flows use randomly selected transmission rates. However, malicious traffic differs significantly in terms of transmission duration and intervals. Specifically, each malicious traffic flow has a randomly assigned transmission duration ranging from 15 to 45 seconds, shorter and distinct from the benign traffic durations. Furthermore, between consecutive malicious flows, random standby intervals ranging from 50 to 100 seconds are implemented. After the completion of all malicious flows in a cycle, another random standby period (again between 50 and 100 seconds) is selected before initiating the next loop. This excess randomness in malicious traffic is added to further emulate real-life attack patterns, where legitimate parties have limited knowledge of an imminent attack \cite{random_attack_ref}. The unpredictability introduced by these varying intervals also simulates realistic operational uncertainty in network security scenarios. The randomness helps to create diverse and challenging datasets. Table \ref{trafficPatterns} shows the details of the malicious traffic.

\begin{table}[t]
\footnotesize
\centering
\caption{Descriptions of Traffic Patterns}
\label{trafficPatterns}
\begin{tabular}{
    >{\centering\arraybackslash}p{0.25\columnwidth} 
    >{\centering\arraybackslash}p{0.16\columnwidth} 
    >{\centering\arraybackslash}p{0.18\columnwidth} 
    >{\centering\arraybackslash}p{0.17\columnwidth}
}
\toprule
\textbf{Type} & \textbf{Protocol} & \textbf{BW (Mbps)} & \textbf{Duration (s)} \\
\midrule
\multicolumn{4}{@{}c}{\textbf{Benign Traffic}} \\ \hline
\addlinespace[0.5ex]
Instant Messaging     & 80 / UDP     & 0.5–1.5 & 85–125 \\
Remote Access         & 22 / TCP     & 1.5–2.5 & 85–125 \\
Web Browsing          & 5222 / TCP   & 1.5–2.5 & 85–125 \\
File Transfer         & 21 / TCP     & 2.5–3.5 & 85–125 \\
Streaming             & 1935 / TCP   & 2.5–3.5 & 85–125 \\ \hline

\addlinespace[0.5ex]
\multicolumn{4}{@{}c}{\textbf{Malicious Traffic}} \\ \hline
\addlinespace[0.5ex]
UDP Flood             & UDP     & 2.5–3.5 & 15–45 \\
SYN Flood             & TCP     & 1.5–2.5 & 15–45 \\
ICMP Flood            & ICMP    & 1.5–2.5 & 15–45 \\
TCP Flood             & TCP     & 1.5–2.5 & 15–45 \\
\bottomrule
\end{tabular}
\vspace{-10pt}
\end{table}

The dataset consists of 103 traces collected over several weeks, corresponding to approximately 10 hours of real-world measurements on a physical testbed. Each sample includes 70 features, of which 69 are KPMs corresponding to three UEs in the network given in Table \ref{kpmMetrics}. More explicitly, the first 23 features correspond to the KPMs of the receiving UE (handling both traffic types) while the final 23 features represent KPMs of the attacker UE, and the remaining features belong to KPMs of the transmitting UE. The final feature is a binary attack flag indicating the presence (1) or absence (0) of malicious traffic at each time step. To enable reproducibility and foster further research, the dataset and corresponding scripts will be made publicly available \vspace{-3 pt}\footnote{{\url{https://github.com/afs-code/XAI-in-WirelessCommunications-EBADF}}}.

\section{Methodology}
\label{methods}
In this section, the methodology of the proposed framework will be discussed. The aim is to provide a comprehensive framework for anomaly detection with various data types and distinctive explainability approaches.   

\subsection{Exponentially Weighted Moving Average}
While post-hoc methodologies can successfully identify the features with the greatest impact for a given decision, they require trained AI/ML models, and their training process can be computationally expensive. To address this issue, statistical-based algorithms can be utilized. Exemplary, an adaptive threshold-based algorithm named the Exponentially Weighted Moving Average (EWMA) has been utilized. While the EWMA method can be employed prior to training of AI/ML models to identify the significant features, this study utilizes the EWMA method solely for explainability purposes. The proposed EWMA method operates as follows:

The time series data observations for the feature $i$ can be denoted by $x_1^i, x_2^i,..., x_t^i,..., x_N^i$ where $t$ and $N$ represent the time instance and total number of observations, respectively. For each time instance, the mean can be denoted as $\mu_t$ while the standard deviation is denoted as $\sigma_t$. Following, upper and lower bounds are denoted as $U_t$ and $L_t$, respectively. In the proposed method, the mean, standard deviation, and bounds change on each time instance with respect to previous observations. The mean for time instance $t$ is updated by a weighted average of the current observation, denoted as $x_t^i$, and with previous mean, denoted as $\mu_{t-1}$, such that,
\begin{equation}
    \mu_t = \alpha_{\mu} x_t^i + (1 - \alpha_{\mu}) \mu_{t-1}, \label{ewma_mean}
\end{equation}
where $\alpha_{\mu}$ is the  hyperparameter for the $\mu_t$. Furthermore, the standard deviation is updated based on the absolute deviation between the current observation and the calculated mean. The deviation at time $t$ can be calculated as, $d_t = |x_t^i-\mu_t|$. Then, the standard deviation at time $t$ is updated as,
\begin{equation}
    \sigma_t = \alpha_{\sigma} d_t + (1 - \alpha_{\sigma}) \sigma_{t-1}, \label{ewma_volatility}
\end{equation}
where $\alpha_{\sigma}$ represent the hyperparameter for $\sigma_{t}$, respectively. Following, the upper, $U_t$, and lower $L_t$ bounds at time $t$ are calculated as, \vspace{-10pt}
\begin{align}
    U_t &= \mu_t + k_{U} \cdot \sigma_t, \label{upper_bound} \\
    L_t &= \mu_t - k_{L} \cdot \sigma_t, \label{lower_bound}
\end{align}
where $k_{U}$ and $k_{L}$  are the multiplier hyperparameters for the upper and lower bounds. Finally, an anomaly, which is denoted as $A_t$, can be detected if it falls outside the dynamic range defined by upper and lower bounds as,
\begin{equation}
A_t =
\begin{cases}
1 & \quad \text{if } x_t^i > U_t \text{ or } x_t^i < L_t. \\
0 & \quad \text{otherwise}.
\end{cases}
\end{equation}
Here, $A_t$ has binary values where $A_t = 1$ corresponds to a detected attack and $A_t = 0$ corresponds to normal behaviour.

\subsection{Artificial Intelligence \& Machine Learning Methods}
ML methods can offer a compelling balance of speed, accuracy, and computational efficiency when compared to purely statistical approaches or more complex AI techniques that may require extensive tuning and higher resource overhead. However, AI techniques can offer more advanced prediction performance at a higher computational cost. Hence, AI/ML models have been chosen to demonstrate the trade-off between these approaches. Additionally, the models are chosen to cover distinctive data types: Decision Tree (DT), Random Forest (RF), and K-Nearest Neighbor (KNN) for tabular and labeled data; Isolation Forest (IF) for tabular and unlabeled data; and lastly, Long Short Term Memory (LSTM) Classifier for sequential and labeled data. 

\subsection{Post-Hoc Methods}
Post-hoc methods will be utilized to estimate the most significant KPMs for detecting anomalies. Post-hoc methods are explainability methods applied after a model has already been trained \cite{postHoc}. Their purpose is to help us understand the reasoning of the model's prediction. These methods are designed for "black-box" models, and they do not require altering the original model structure or the training process. Instead, they analyze the trained model’s behavior, often by examining how changes to the input features affect its outputs. With this ability, decision-makers can gain insight into which features drive predictions and how sensitive the model is to different factors, improving transparency and trust in the results. In this study, two post-hoc methods are employed: Local Interpretable Model-agnostic Explanations (LIME) \cite{lime} and SHapley Additive exPlanations (SHAP) \cite{shap}. While LIME approximates the model into a simpler, linear model for the given sample, SHAP calculates "Shapley values" to identify the contribution of each feature. While this study only utilizes the EWMA method for local explainability to identify critical attack characteristics, global explainability results have been investigated through EWMA, LIME, and SHAP methods to reduce the overhead for the data collection process.

\section{Results}
\label{results}

\begin{table*}[t]
\vspace{5pt}
\centering
\caption{Comparison of Selected Methods on All vs. Significant Features}
\vspace{-5pt}
\label{allResults}
\resizebox{\textwidth}{!}{%
\begin{tabular}{|c|c|c|c|c|c|c|c|c|}
\hline
\textbf{\!Features\!} & \textbf{\!\!Method\!\!}&
 \textbf{\!\!Train Acc. \!\!(\%)\!\!} & \textbf{\!\!Test Acc. \!(\%)\!\!} & \textbf{\!\!Train Time \!(s)\!\!} & \textbf{Inf. Lat. ($\mu$s)} & \textbf{\!Precision\!} & \textbf{\!Recall\!} & \textbf{\!F1-Score\!} \\
\hline
\multirow{5}{*}{All Features} & Decision Tree (DT) & 97.04 & 96.25 & 0.1881 & \textbf{1.60} & 0.957 & 0.922 & 0.939   \\
& \textbf{Random Forest (RF)} & \textbf{97.42} & \textbf{97.01} & 0.2132 & 24.70 & \textbf{0.973} & \textbf{0.930} & \textbf{0.951} \\  
& K-Nearest Neighbor (KNN) & 94.70 & 90.80 & \textbf{0.0027}  & 17.70 & 0.861 & 0.842 & 0.852  \\
& Isolation Forest (IF) &  80.95 & --- & 0.2416 & 99.00 & 0.709 & 0.679 & 0.694 \\ 
& Long Short Term Memory (LSTM)\!\! &  95.82 & 94.70  & 167.8298 & 25.60 & 0.924 & 0.895 & 0.909  \\ \hline 
\multirow{5}{*}{Significant Features} & Decision Tree (DT) & 97.00 & 96.75 & 0.0484 & \textbf{0.91} & \textbf{0.973} & 0.922 & 0.947      \\ 
& \textbf{Random Forest (RF)} & \textbf{97.38} & \textbf{96.99} & 0.1655 & 25.20 & \textbf{0.973} & \textbf{0.928} & \textbf{0.950} \\ 
& K-Nearest Neighbor (KNN) & 97.34 & 96.20 & \textbf{0.0102}  & 1.89 & 0.968 & 0.909 & 0.937  \\
& Isolation Forest (IF) &  88.69 & ---  & 0.2214 & 86.09 & 0.832 & 0.797 & 0.814 \\ 
& Long Short Term Memory (LSTM) \!\! &  96.39 & 96.33  & 47.6151 & 20.32 & 0.971 & 0.907 & 0.938  \\ \hline
\end{tabular}
}
\vspace{-10pt}
\end{table*}

For the anomaly detection analysis, the first step includes the partition of the anomaly detection dataset into training and test subsets using an 80–20 split, yielding 8000 training and 2000 test samples. The IF model is the sole exception: it is trained on the entire dataset to emulate an unsupervised learning application. Secondly, model parameters have been selected via the grid search approach. To promote robustness and reproducibility, ten training runs per model are conducted with fixed random seeds. Table \ref{allResults} presents the accuracy performances on both datasets, alongside runtime information. For comprehensive analysis, precision, recall, and F1-score metrics have been investigated as well. 

\begin{table}[t]
\centering
\caption{Significant Features}
\vspace{-5pt}
\label{tab:features_updated}
\begin{tabular}
{
    >{\centering\arraybackslash}p{0.52\columnwidth} 
    >{\centering\arraybackslash}p{0.15\columnwidth} 
    >{\centering\arraybackslash}p{0.17\columnwidth} 
}
\hline
\hspace{-7pt} \textbf{Feature} & \hspace{-10pt} \textbf{Method} & \hspace{-10pt} \textbf{EWMA Rank} \\
\hline
\hspace{-7pt} \textbf{AttackerUE\_DRB.UEThpUl} & \hspace{-10pt} Both (1/1) & \hspace{-10pt} 2 (0.0275) \\
\hspace{-7pt} \textbf{AttackerUE\_DRB.PdcpSduVolumeUL} & \hspace{-10pt} Both (2/2) & \hspace{-10pt} 1 (0.0276) \\
\hspace{-7pt} AttackerUE\_DRB.UEBufferSizeUL &  \hspace{-10pt} LIME (3) & \hspace{-10pt} 8 (0.0245) \\
\hspace{-7pt} AttackerUE\_RRU.PrbTotDL     & \hspace{-10pt} SHAP (3) & \hspace{-10pt} 9 (0.0235) \\
\hline
\end{tabular}
\vspace{-5 pt}
\label{significantFeatures}
\end{table}

\begin{figure}[t]
    \begin{subfigure}{0.49\columnwidth}
        \centering
        \includegraphics[width=\columnwidth]{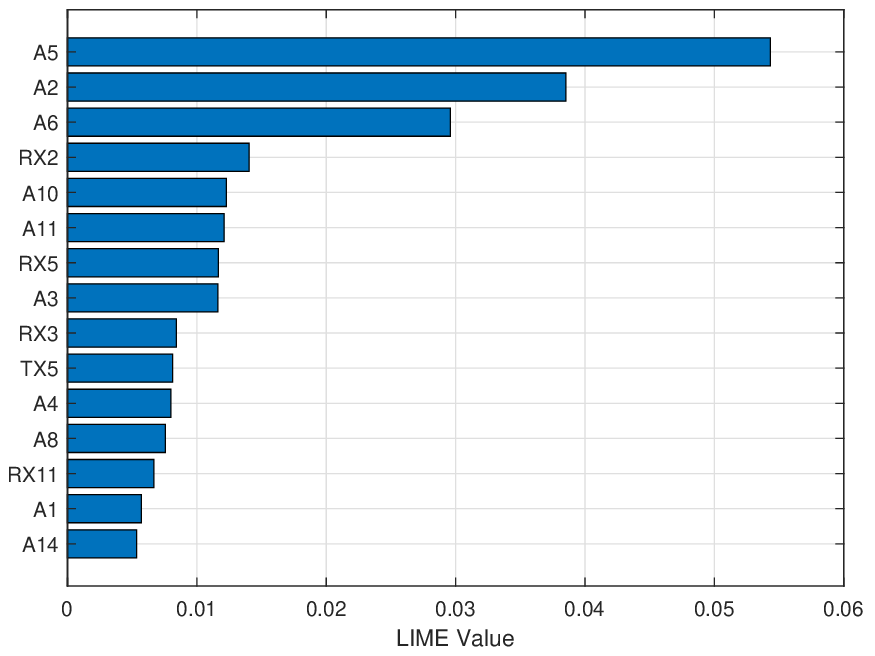}
        \vspace{-10 pt}
        \caption{LIME}
        \label{limeResults}
    \end{subfigure}
        \begin{subfigure}{0.49\columnwidth}
        \centering
        \includegraphics[width=\columnwidth]{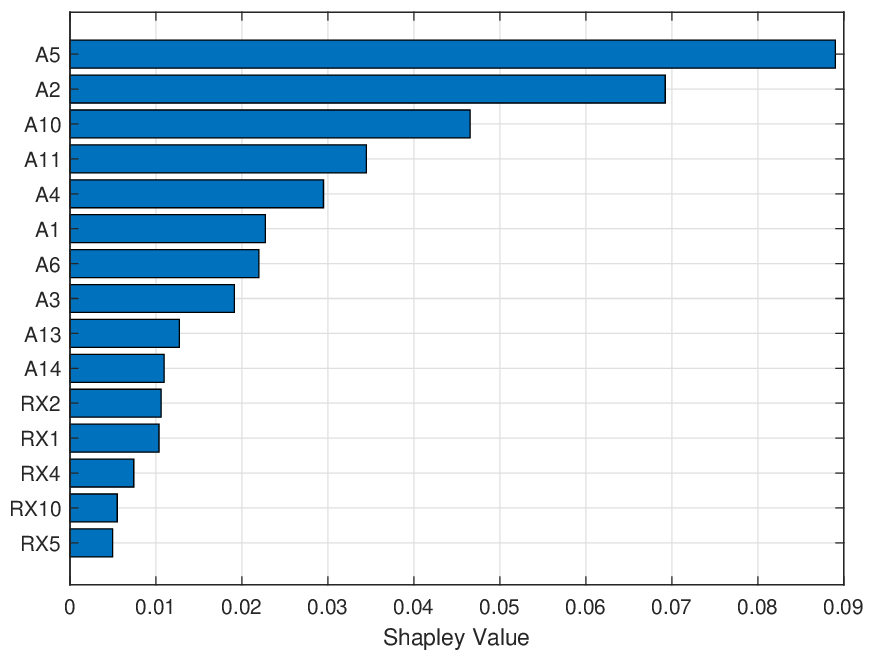}
        \vspace{-10 pt}
        \caption{SHAP}
        \label{shapResults}
    \end{subfigure} \\ 
    \begin{subfigure}{0.99\columnwidth}
        \centering
        \includegraphics[width=\columnwidth]{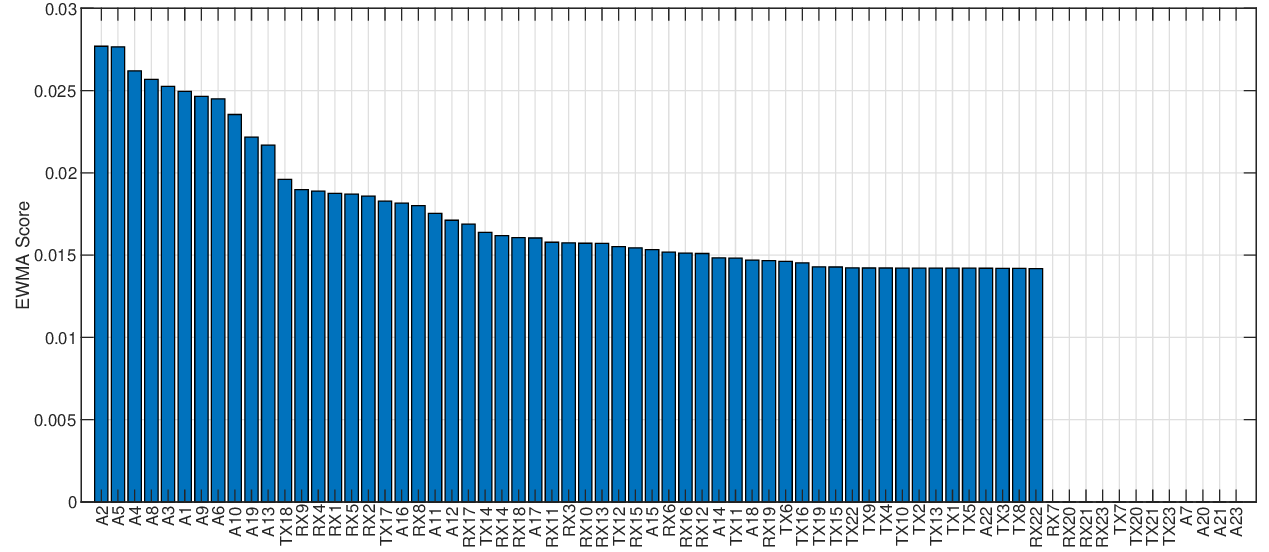}
        \vspace{-10 pt}
        \caption{EWMA}
        \label{ewmaResults}
    \end{subfigure}
    \caption{Global explainability analysis.}
    \vspace{-10 pt}
\end{figure}

Results show high training accuracy for four out of five models, with the only exception being IF. This is consistent with its design for unsupervised anomaly detection. On both training and test datasets, RF attains the highest accuracy performance, followed by DT. The only overfitting tendency is observed for KNN, while the remaining models can generalize well. In terms of computational efficiency, KNN achieves the best runtime in our setting due to its minimal training cost. On the contrary, LSTM has a significantly higher computational cost than the remaining models due to the large number of trainable parameters present. Notably, DT achieves the lowest inference latency among all models, making it particularly well suited for near-RT deployment scenarios. Consistent with accuracy performance, RF delivers the strongest precision, recall, and F1-score performance. DT comes close second, and LSTM has the third-best performance. While KNN trails these three models, IF again underperforms on these metrics.

To obtain consistent and factual explainability results, the model must have a great ability to detect anomalies to lead to accurate explanations. Hence, the RF model has been chosen for explainability analysis due to its superior performance. Following the selection of the model, LIME and SHAP methods have been implemented. Corresponding results can be seen in Fig. \ref{limeResults} and Fig. \ref{shapResults}, respectively. Here, scores have been normalized for fair assessment. To reduce the overhead, a filtering approach has been implemented. Since top features have significantly higher scores, the filtering boundary has been selected as the \%50 of the score for the most impactful feature. Additionally, the notation is as follows: $TX$, $RX$, and $A$ represent the transmitter, receiver, and attacker UEs given in Fig. \ref{fig:systemModel}, whereas the number represents the KPM index given in Table \ref{kpmMetrics}, respectively. 

For LIME, the most influential feature is \textit{A\_DRB.UEThpUl}, followed by \textit{A\_DRB.PdcpSduVolumeUL}. Both features reflect uplink data activity and are consistent with the expected attack pattern. The SHAP analysis yields the same top two features, further corroborating this behavior. The third-ranked feature differs for both methods: LIME highlights \textit{A\_DRB.UEBufferSizeUL}, indicative of flooding where the attacker saturates buffers to deny service to other UEs, whereas SHAP identifies \textit{A\_RRU.PrbTotDL}, suggestive of error-check–driven protocol attacks (e.g., TCP/ICMP) that induce redundant transmissions by victim UEs. Applying the predefined filtering rule, the significant features in Table \ref{significantFeatures} are selected to form the Significant Features dataset. Analogously, EWMA is employed to validate the post-hoc findings. Figure \ref{ewmaResults} reports normalized F1-scores through the EWMA method (EWMA scores) for all features. As expected, all significant features rank highly; the top two again appear as the two most impactful. The remaining significant features also achieve strong standings, appearing at ranks 8 and 9, respectively.

After identifying the significant features, a second training phase is conducted on the Significant Features dataset. Because the attacker cannot be known a priori, a total of 12 features are retained to emulate realistic conditions. For consistency, model hyperparameters have been kept the same. The results in Table \ref{allResults} show that RF and DT maintain strong accuracy despite the reduced feature set, while the remaining models improve, and KNN no longer overfits. Runtime decreases for RF, DT, and LSTM; IF remains similar, and KNN becomes slightly slower. Consistent with the previous setting, DT again achieves the fastest inference latency. Moreover, KNN benefits substantially from the reduced feature set, yielding inference latency comparable to DT. RF preserves its performance, and all other models improve with IF exhibiting the largest gains. These outcomes indicate that post-hoc feature selection can simultaneously reduce overhead and enhance learning.

\begin{figure}[t]
    \centering
    \includegraphics[width=0.5\textwidth]{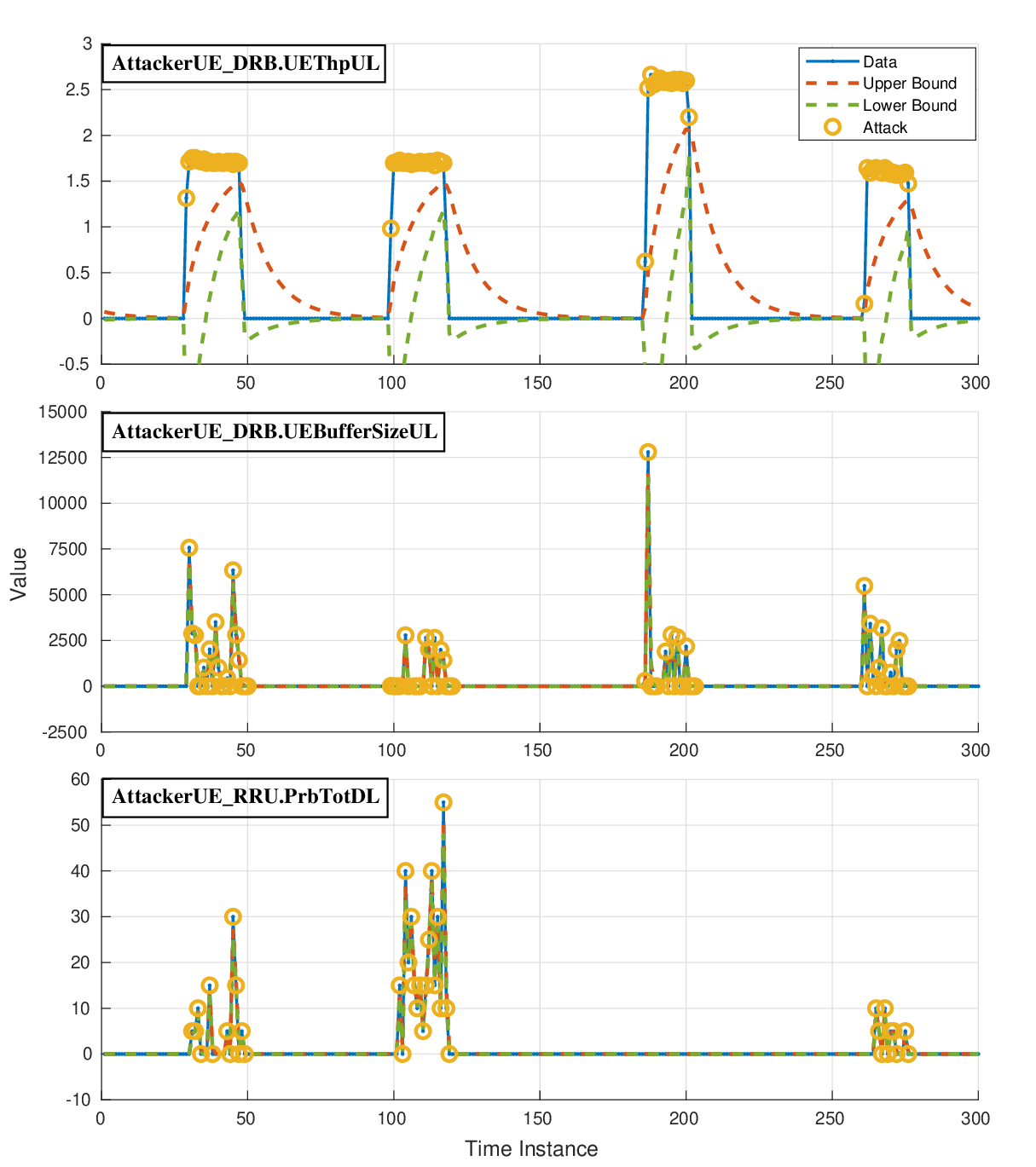}
    \caption{Sequence-based explainability analysis.}
    \label{ewma_AD}
    \vspace{-15 pt}
\end{figure}

Figure \ref{ewma_AD} shows the local explainability analysis for significant features with the EWMA method. Because \textit{A\_DRB.UEThpUl} and \textit{A\_DRB.PdcpSduVolumeUL} exhibit nearly identical temporal patterns; only the former has been reported. For \textit{A\_DRB.UEThpUl}, EWMA reliably detects anomalies under randomized bandwidths, attack intervals, and durations, consistent with the setup in Section \ref{SectionIII}. Similarly, \textit{A\_DRB.UEBufferSizeUL} and \textit{A\_RRU.PrbTotDL} metrics demonstrate the difference in attack protocol types, as the third attack seems to be a UDP attack, while the rest seem to be error-check-based protocol attacks due to the increase in the resource blocks for the downlink direction.

\section{Conclusion}
\label{conclusion}
In this study, an O-RAN-based anomaly detection framework leveraging AI/ML models is proposed. Dedicated traffic generation tools are employed to simulate realistic benign and malicious traffic scenarios. Unlike legacy datasets such as NSL-KDD and 5G-NIDD, our dataset consists of real-time O-RAN standardized performance metrics with randomized patterns. Among the tested models, RF demonstrates superior performance. Post-hoc explainability identifies critical KPMs for anomaly detection. By leveraging the identified features, the complexity of the dataset is reduced by approximately 80\%, without compromising the performance. Explainability analyses show the critical attack characteristics for the prevention of upcoming attacks. Finally, inference latency analysis confirms that the proposed mechanism is suitable for near-RT applications. Future work will explore unsupervised anomaly detection applications, incorporate detection methods for zero-day attacks, and investigate O-RAN-based latency and overhead analysis for real-time deployment of the proposed framework.

\vspace{-5 pt}
\bibliographystyle{IEEEtran}
\bibliography{refer} 

\end{document}